\documentclass[onecollarge,natbib]{svjour2}
\bibpunct{[}{]}{;}{n}{}{,} 
\smartqed  
\usepackage{graphicx}
\usepackage{amsmath}
\journalname{Few Body Systems}
\begin{document}
\title{Infra-red divergences in Light Front Field Theory and Coherent State Formalism}

\author{Jai D. More \and Anuradha Misra}
\institute{Jai D. More \at 
              Department of Physics, University of Mumbai, SantaCruz(E), Mumbai, India-400098. \\
              \email{more.physics@gmail.com}           
           \and
          Anuradha Misra\at
              Department of Physics, University of Mumbai, SantaCruz(E), Mumbai, India-400098.\\
              \email{anuradha.misra@gmail.com}
}
\date{Received: date / Accepted: date}
\maketitle
\begin{abstract}
We discuss fermion self energy correction in light front QED using a coherent 
state basis. We show that if one uses coherent state basis instead of fock basis to calculate the 
transition matrix elements the true infrared divergences in $\delta m^2$ get canceled up to $O(e^4)$ . We show this 
in Light-front as well as in Feynman gauge.
\keywords{Light Front QED \and Coherent State \and Improved Method of Asymptotic dynamics}
\end{abstract}
\section{INTRODUCTION}\label{intro}

Our notation for light front coordinates \cite{MUS91} is 
 \begin{displaymath}
 x^{\mu}=(x^+,x^-,{\bf x}^{\perp}) 
\end{displaymath}
where
\begin{displaymath}
x^+=\frac{(x^0 +x^3)}{\sqrt{2}},\quad x^-=\frac{(x^0 -x^3)}{\sqrt{2}}, \quad {\bf x}_{\perp}=(x^1,x^2)
\end{displaymath}
Momentum is given by
\begin{displaymath}
p^\mu=(p^+, p^-, {\bf p}_\perp) 
\end{displaymath}
Mass shell condition is  
\begin{displaymath}
  p^-=\frac{p^2_\perp+m^2}{2p^+}
\end{displaymath}
and the metric tensor is 
\[
g^{\mu\nu} =
\left[{\begin{array}{cccc}
0&1&0&0\\
1&0&0&0\\
0&0&-1&0\\
0&0&0&-1\\
\end{array}}\right]
\]

In quantum field theory, the LSZ formalism is based on the assumption that at large times, the dynamics of incoming and outgoing particles in a scattering process is governed by the free Hamiltonian
\begin{equation}
H_{as}=\lim_{\left |t\right| \rightarrow \infty} H= H_0
\end{equation}
However, it was pointed out by Kulish and Faddeev \cite{KUL70} that this assumption does not hold for theories in which either long range interactions like QED are present or the incoming and outgoing states are bound states like QCD. Thus in large time limit, the free Hamiltonian should be replaced by a new Hamiltonian that must contain interaction terms in such a way that all soft singularities of full Hamiltonian are included. 
This new Hamiltonian is called asymptotic Hamiltonian. Kulish and Faddeev (KF) proposed the method of asymptotic dynamics and showed that in QED, at large times, when one takes into account the long range interaction between the incoming and outgoing states, then $H_0 \neq H_{as}$. 
Using the method of asymptotic dynamics, interaction Hamiltonian $V_{as}$ in 
\begin{equation}
H_{as}=H_0 +V_{as}
\end{equation}
was shown by KF to be non-zero in QED. It was then used to construct the asymptotic M\"oller operators
\begin{equation}
\Omega_{\pm}^A = T~exp\biggl[ -i \int^0_{\mp} V_{as}(t)dt \biggr]
\end{equation}
which leads to the coherent states
\begin{equation}
\vert n \colon coh \rangle = \Omega_{\pm}^A \vert{n}\rangle \;,
\end{equation}
If one computes the transition matrix element using these coherent states obtained from KF approach, the IR divergence are cancelled. 

There are two kinds of IR divergences in LFFT:
\begin{itemize}
\item[1)] {\bf Spurious IR divergences}: These divergences arise when $k^+\rightarrow 0$ and are actually a manifestation of UV divergences of equal time theory. These are regularized by an infrared cut-off on small values of longitudinal momentum.
\item[2)] {\bf True IR divergences}: These are the actual IR divergences of equal time theory and are present because of the particle being on mass shell. There are different ways of handling these divergences, for example using mass regularization. An alternative treatment to this problem is provided by coherent state method.
\end{itemize}

A coherent state approach based on asymptotic dynamics has been developed for LFFT and has been applied to lowest order calculations 
\cite{ANU94, ANU96, ANU00}. In this approach, $H_{as}$ is evaluated by taking the limit $x^+\rightarrow \infty$ 
in $exp[-i(p_1^-+p_2^-+\cdots+p_n^-)x^+]$ which contains the light-cone time dependence of the interaction Hamiltonian 
$H_{int}$. If $(p_1^-+p_2^-+\cdots+p_n^-)\rightarrow 0$ for some vertex, then the corresponding term in 
$H_{int}$ does not vanish in large $x^+$ limit. One can then use KF method to obtain the asymptotic Hamiltonian which can then be used to construct the asymptotic M\"oller operator and coherent states. True IR divergences are not expected to appear when one uses these coherent states to calculate the transition matrix elements.

Interaction Hamiltonian of LFQED in LF gauge is given by 
\begin{displaymath}
H_I(x^+)=V_1(x^+)+V_2(x^+)+V_3(x^+)
\end{displaymath}
where
\begin{equation}\label{V_1}
V_1(x^+)= e\sum_{i=1}^4 \int d\nu_i^{(1)}[ e^{-i \nu_i^{(1)} x^+} {\tilde h}_i^{(1)}(\nu_i^{(1)})
+ e^{i \nu_i^{(1)} x^+}
{\tilde h}^{(1)\dagger}_i (\nu_i^{(1)})]
\end{equation}
${\tilde h}^{(1)}_i(\nu^{(1)}_i)$ and $\nu_i^{(1)}$ are three point QED interaction vertex and the light-front energy transferred at the vertex ${\tilde h}^{(1)}$ respectively. 
$V_2$ and $V_3$ are the non-local 4$-$point instantaneous vertices.
Here, we will focus on construction of asymptotic Hamiltonian using 3$-$point vertex. 
One can notice, from the time dependence, that $V_1(x^+)$ does not actually become zero at large times if 
$\nu_i^{(1)}=p^--k^--(p-k)^-= 0$. We define the asymptotic region (i.e the region in which the light-cone energy difference in the exponent in Eq.~\ref{V_1} is zero) as 
\begin{displaymath}
{\bf k}_\perp ^2 < {{k^+ \Delta} \over{p^+}} \quad k^+ < {{p^+ \Delta} \over {m^2}}\;.
\end{displaymath}
Thus, the asymptotic Hamiltonian is given by
\begin{eqnarray}
V_{1as}(x^+) = e \sum_{i=1,4} \int d\nu_i^{(1)} \Theta_\Delta(k)[ e^{-i \nu_i^{(1)} x^+} \tilde h_i^{(1)}(\nu_i^{(1)})
+ e^{i \nu_i^{(1)} x^+}\tilde h^ {\dagger}_i (\nu_i^{(1)})] \;
\end{eqnarray}
where $\Theta_\Delta(k)$ is given by
$$ \Theta_\Delta(k)=\theta\bigg({{k^+\Delta} \over p^+} - {\bf k}_\perp^2\bigg)
\theta\bigg({{p^+\Delta} \over m^2} - k^+\bigg)$$

Substituting $k^+\rightarrow 0$, $k_\perp \rightarrow 0 $ in all slowly varying functions of k and performing the 
$x^+$ integration one obtains the asymptotic M\"oller operator which gives the asymptotic states as
\begin{align}
\Omega_{\pm}^A \vert n\colon p_i \rangle=&exp\biggl[-e\int{dp^+d^2{\bf p}_\perp}\sum_{\lambda=1,2}[d^3k][f(k,\lambda:p) 
a^\dagger(k,\lambda)- f^*(k,\lambda:p)a(k,\lambda)]\nonumber\\
&+e^2\int{dp^+d^2{\bf p}_\perp} \sum_{\lambda_1, \lambda_2=1,2}[d^3k_1][d^3k_2][g_1(k_1,k_2,\lambda_1,\lambda_2 \colon p) 
a^\dagger(k_2,\lambda_2)a(k_1,\lambda_1)-\nonumber\\
&g_2(k_1,k_2,\lambda_1,\lambda_2 \colon p)a(k_2,\lambda_2)a^\dagger(k_1,\lambda_1)]
\rho(p)\biggr]\vert n \colon p_i \rangle
\end{align}
The second term here arises from the 4$-$point instantaneous interaction. In our work \cite{JAI12}, we have used these asymptotic states to calculate the transition matrix elements and to demonstrate the absence of IR divergences in them.

The light-front QED Hamiltonian in the light-front gauge consists of the free part, the standard three point QED vertex and two 4$-$point instantaneous interactions,
\begin{displaymath}
P^-= H \equiv H_0 + V_1 + V_2 + V_3 \;,
\end{displaymath}
Here,
\begin{align}
H_0=&\int d^2 {\bf x}_\perp dx^- \{\frac{i}{2} \bar{\xi}\gamma^-
\stackrel{\leftrightarrow}{\partial}_-\xi + \frac{1}{ 2} (F_{12})^2-\frac{1}{2}a_+\partial_- \partial_k a_k \}\\
V_1=&e \int d^2{\bf x}_\perp dx^- \bar{\xi} \gamma^{\mu}\xi a_\mu\
\end{align}
$V_2$ and $V_3$ are the 4$-$point instantaneous interaction terms \cite{JAI12}. $\xi(x)$ and $a_\mu(x)$ can be expanded in terms of creation and annihilation operators \cite{MUS91}.
In LFFT, usually one uses light front time ordered perturbation theory to calculate Hamiltonian matrix elements. The transition matrix is given by the perturbative expansion 
\begin{displaymath}
T= V + V {1 \over {p^--H_0}}V + \cdots
\end{displaymath}
The electron mass shift is obtained by calculating the matrix element of this series ($T_{pp}$ ) between the initial and final electron states $\vert p,s \rangle$. One can expand $T_{pp}$ in powers of $e^2$ as 
\begin{equation}
T_{pp}=T^{(1)}+T^{(2)}+\cdots
\end{equation}
where $T^{(n)} $ gives the $O(e^{2n})$ contribution to fermion self-energy correction. 
\section{Mass renormalization up to $O(e^2)$}
$O(e^{2})$ correction is represented by two diagrams as shown in Fig. 1. Since the second diagram is a tree level diagram and does not have any vanishing denominator, we need to calculate only the first diagram yielding
\begin{align}
T^{(1)}_{pp}\equiv T^{(1)}( p,p)=\langle p,s \vert V_1 \frac{1} {p^- - H_0}V_1\vert p,s \rangle
\end{align}
We calculate this using the procedure in Ref.~\cite{JAI12} and obtain $\delta m^2_{1a}$ which reduces 
in the limit $k_1^+ \rightarrow 0$, ${\bf k}_{1\perp} \rightarrow 0$ to 
\begin{eqnarray}\label{deltam1}
\boxed{
{(\delta m^2_{1a})}^{IR}=-\frac{e^2}{(2\pi)^3}\int{d^2{\bf k}_{1\perp}}\int\frac{dk_1^+}{k_1^+}\frac{(p\cdot \epsilon(k_1))^2}{(p\cdot k_1)}
}
\end{eqnarray}

It should be noted that this is only the IR divergent part of $\delta m^2_{1a}$. The full expression has more terms in the numerator which reduces to the numerator in Eq.~(\ref{deltam1}) as $k^+, {\bf k}_\perp \rightarrow 0$. This result was derived using LF gauge. 
We have also done our calculations in Feynman gauge.
\begin{figure}[h]
\begin{minipage}{.5\textwidth}
 \centering
\includegraphics[width=0.8\linewidth]{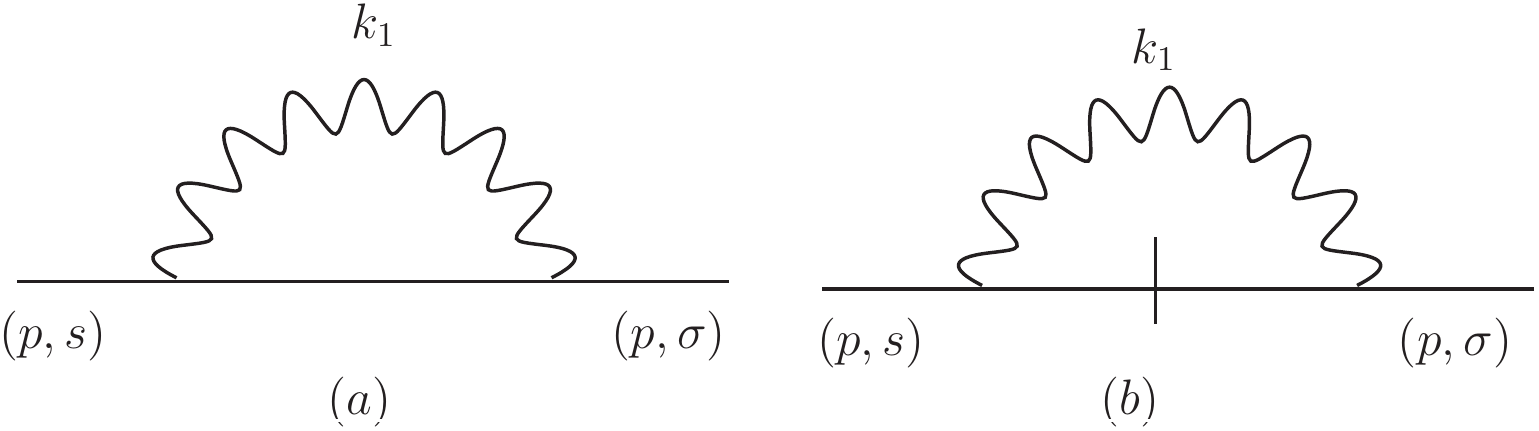}
\caption{Diagrams for $O(e^2)$ self-energy correction in LF gauge.}
\end{minipage}%
\quad
\begin{minipage}{.5\textwidth}
\centering
  \includegraphics[width=0.8\linewidth]{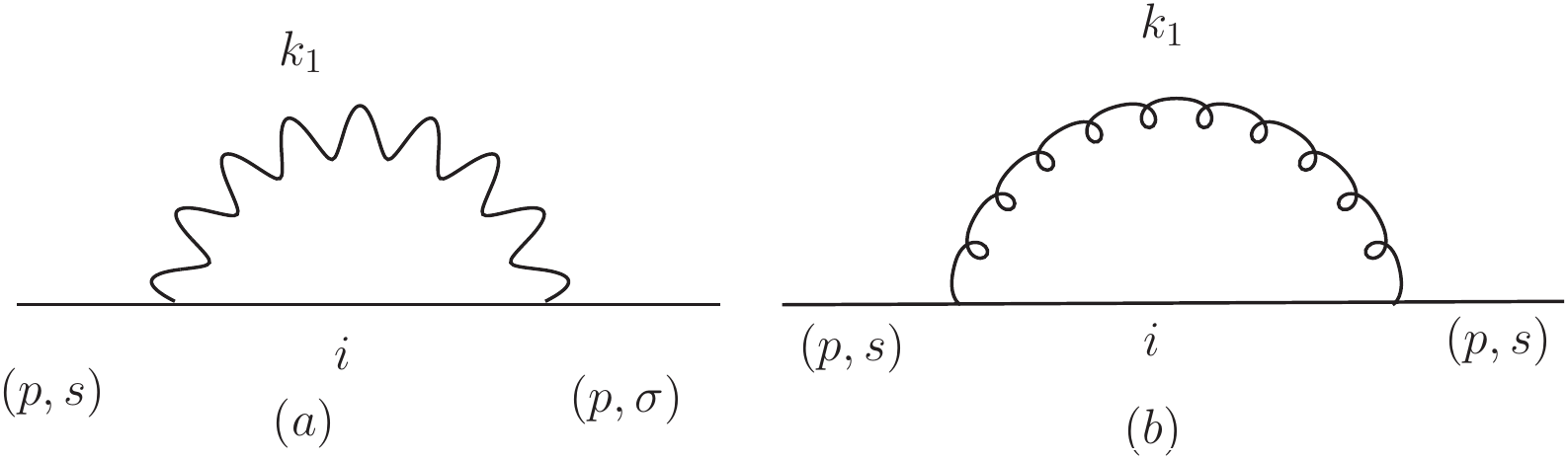}
\caption{The contributions to $O(e^2)$ self-energy correction in Feynman gauge.}
\end{minipage}
\end{figure}
QED Lagrangian in Feynman gauge with additional PV fields is given by \cite{HILLER11}: 
\begin{eqnarray}  
{\cal L}=\sum_{i=0}^2 (-1)^i \left[-\frac14 F_i^{\mu \nu} F_{i,\mu \nu} 
         +\frac{1}{2} \mu_i^2 A_i^\mu A_{i\mu} 
         -\frac{1}{2} \left(\partial^\mu A_{i\mu}\right)^2\right] + \sum_{i=0}^2 (-1)^i \bar{\psi_i} (i \gamma^\mu \partial_\mu - m_i) \psi_i-e\bar{\psi}\gamma^\mu \psi A_\mu.
\end{eqnarray}
  
In this theory, there is additional contribution to self-energy correction due to the second diagram in Fig. 2 where the curly line denotes the massive PV field. The index i on the internal fermion line takes values $i = 0$, 1 or
2, where $i = 0$ corresponds to the physical fermion field while ($i = 1$ and 2) represents PV
fermion fields. Note further that there are no instantaneous diagrams now as the non-local terms in the Hamiltonian get cancelled due to PV fields. In addition, for massive photon there is no vanishing denominator for Fig. 2(b) and thus it does not contribute to IR divergences. Therefore, at this order, in both LF and Feynman gauge, there is only one diagram that can give IR divergences. 
In coherent state basis, there are additional contributions in addition to those already discussed for Fock basis. In particular, in $O(e^2)$, one can also get a contribution from diagrams in Fig. 3 
because the coherent state in 
\begin{figure}[h]
\centering
\includegraphics[scale=0.4]{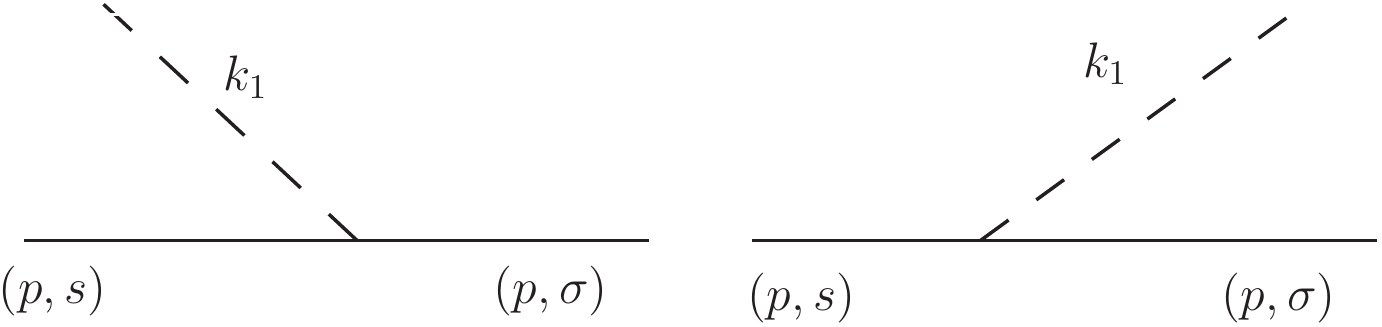}
\caption{Additional diagrams in coherent state basis for $O(e^2)$ self energy correction corresponding to $T_2$}
\end{figure}
\begin{equation}
T^\prime(p,p)=\langle p,s \colon f(p) \vert V_1\vert p,s \colon f(p) \rangle
\end{equation}
contains $O(e)$ terms. In the diagrammatic representation of these contributions given in Fig. 3, the first 
diagram represents a situation where a soft photon accompanying the incoming particle is absorbed and 
the second diagram represents a situation where a soft photon is emitted which accompanies the outgoing 
electron but the two particle states are indistinguishable from the single particle state. 

These coherent state diagrams give a contribution 
\begin{align}
T^\prime(p,p)=\frac{e^2}{(2\pi)^{3}}\int\frac{d^2{\bf k}_{1\perp}}{2p^+}\int \frac{dk_1^+}{2k_1^+}\overline u (\overline p,s^\prime) \epsilon\llap/^\lambda(k_1)u(p,s)f(k_1,\lambda:p)
\end{align}
where $f(k,\lambda \colon p) $ ensures that the integrals are performed only over a small region around $k^+=0$, ${\bf k}_\perp=0$. 
In this region, it gives a contribution equal and opposite to the one loop self-energy correction calculated earlier and thus 
cancels the IR divergent contribution arising due to $p.k_1 \rightarrow 0$ in Eq.~(\ref{deltam1}).
\section{Mass renormalization up to $O(e^4)$}

At $O(e^4)$, self-energy correction is obtained by evaluating
\begin{eqnarray}
T^{(2)}=T_{3}+T_{4}+T_{5}+T_{6}+T_{7}                       
\end{eqnarray}
where $T_3$ contains only 3-point vertices, $T_4$, $ T_5 $ and $ T_6$ contain both 3-point and 
4-point vertices, while $T_7$ contains only 4-point vertices. We have calculated these diagrams using light cone time ordered perturbation theory and have shown that there are IR divergences when either $k_1$ approaches zero or $k_2 $ approaches zero or both approach zero. Details can be found in \cite{JAI12}. 
In Feynman gauge calculation, we do not have the diagrams involving 4-point interactions but there are additional diagrams as shown in Fig 4. 
\begin{figure}[h]
\begin{minipage}{.5\textwidth}
 \centering
\includegraphics[width=0.8\linewidth]{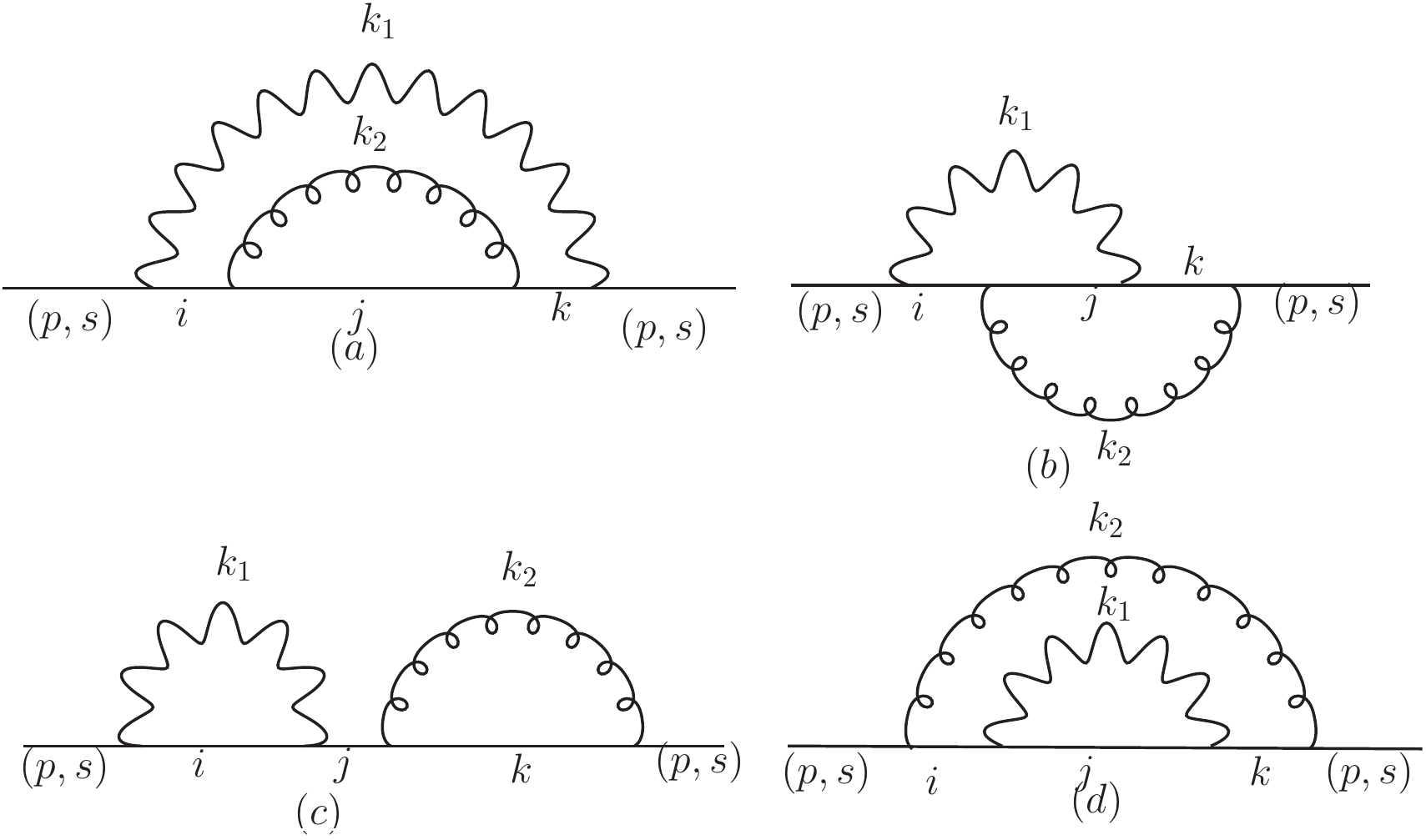}
\caption{Additional diagrams for $O(e^2)$ self energy correction in fock basis in Feynman gauge}
\end{minipage}%
\quad
\begin{minipage}{.5\textwidth}
\centering
  \includegraphics[width=0.35\linewidth]{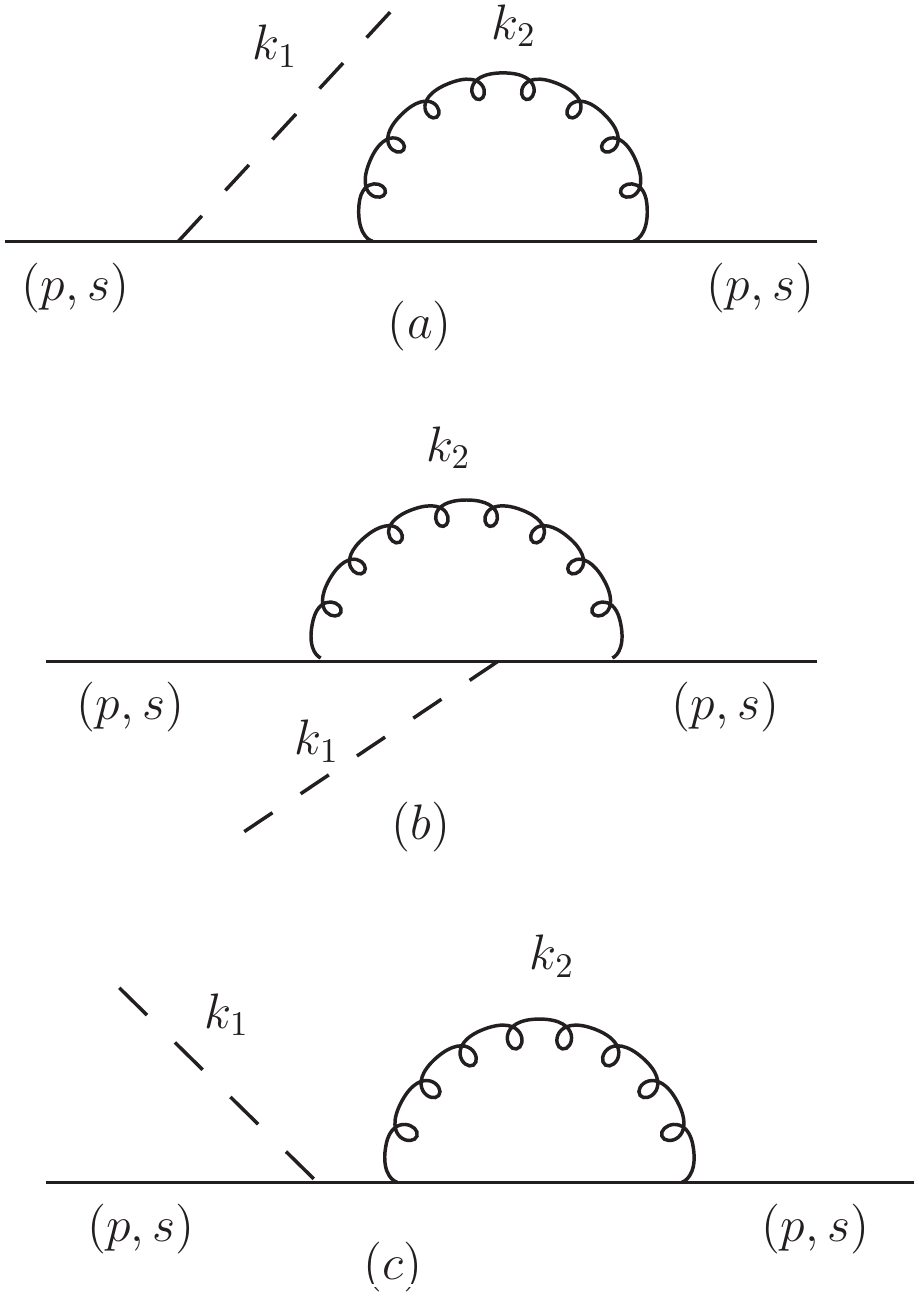}
\caption{Additional diagrams in coherent state basis for $O(e^4)$ self energy correction in Feynman gauge}
\end{minipage}
\end{figure}
Just as in $O(e^2)$, when we add all these diagrams to the coherent state basis diagrams, all the terms involving vanishing energy denominators cancel in the IR region. 
However when one goes to $O(e^4)$, one finds different sets of IR divergent diagrams in the two gauges. In LF gauge, there are diagrams involving instantaneous vertices whereas in Feynman gauge there are diagrams involving PV fermion fields and massless photon. 
However, both the formulations are equivalent due to the simple observation that the four point instantaneous terms can be 
obtained in the infinite PV mass limit. Diagrammatically we can see in Figs. 6 and 7 that PV fermion line reduces to an instantaneous 4$-$point interaction term \cite{HILLER11}.
\begin{figure}[h]
\begin{minipage}{.5\textwidth}
 \centering
  \includegraphics[width=0.85\linewidth]{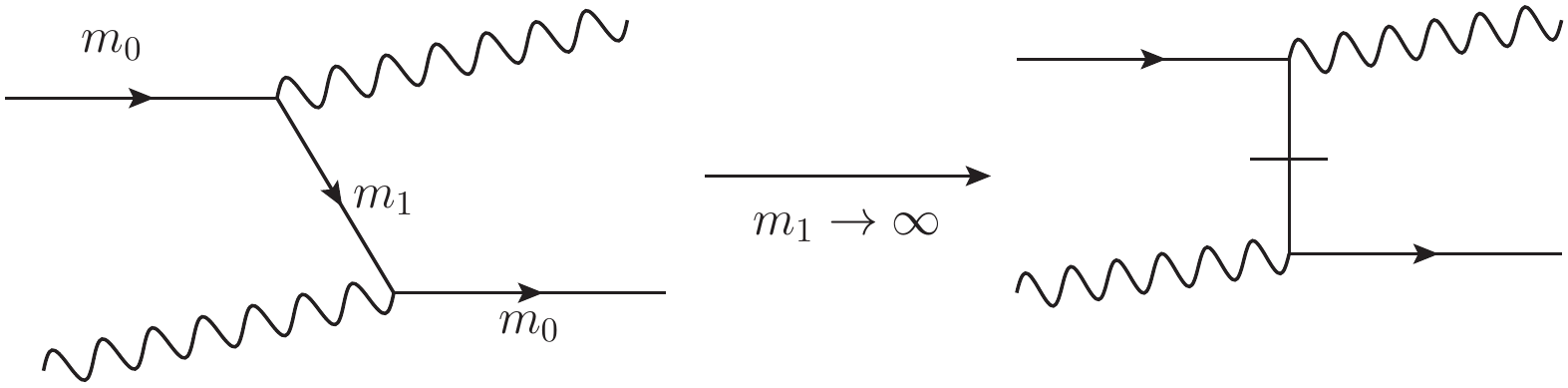}
  \caption{In the infinite PV mass limit the PV fermion line reduces to as instantaneous four-point interaction term denoted by a dash on fermion line}
\end{minipage}
\quad
\begin{minipage}{.5\textwidth}
   \centering
  \includegraphics[width=0.85\linewidth]{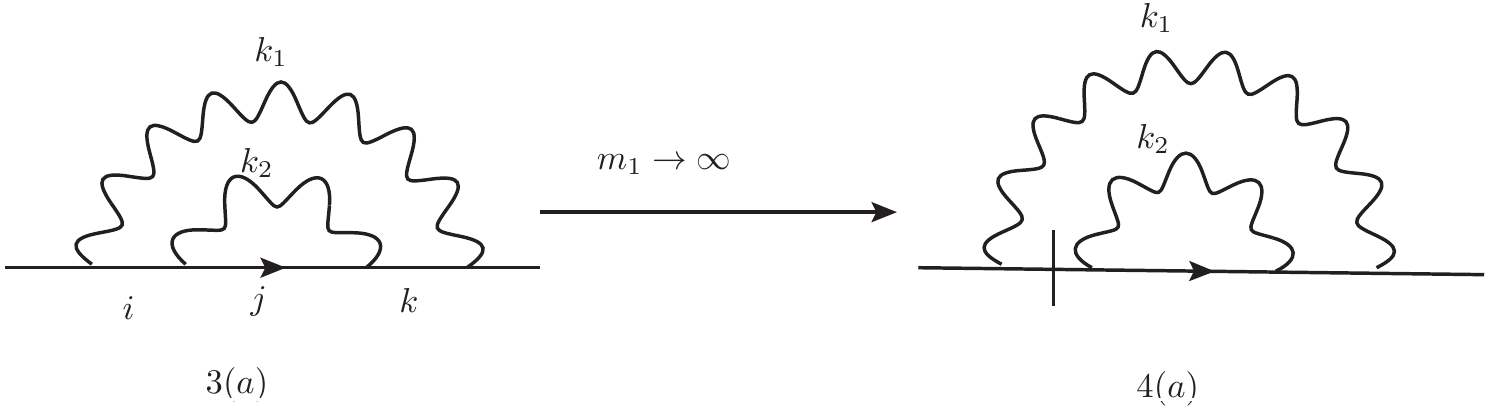}
  \caption{In the infinite PV mass limit the diagram on the left reduces to a diagram involving instantaneous interaction. Here $i = 1$ while $j=k=0$.}
\end{minipage}
\end{figure}
It is very interesting to note that in Fock basis, diagrams containing only massless physical photons have divergences only when $k_1 \rightarrow 0$. The second limit $(k_2 \rightarrow 0)$ and the double limit $(k_1 \rightarrow 0, k_2 \rightarrow 0)$ 
do not give any IR divergences here. Fig. 5 represents the complete set of additional diagrams in coherent state basis at $ O(e^4)$. 
\section{Improved Method of Asymptotic Dynamics}
In equal time formulation of QED, it has been shown using coherent states method that the IR divergences cancel to all orders. In QCD such a proof does not exist. The reason lies in the fact that in QCD, the asymptotic states are bound states. So the asymptotic Hamiltonian obtained by KF method is not sufficient for the cancellation of IR divergences. 

A new approach to the asymptotic dynamics has been proposed called the "improved method of asymptotic dynamics", based on a proposal
by McMullan etal \cite{MUL98,MUL99,MUL00}, that the KF method makes no connection between large time limit and the separation of particles at large 
distances. In particular, they showed that the KF method does not lead to correct result for $\phi^4$ theory, and pointed out 
that it is more appropriate in QFT to work at the level of matrix elements than at the level of operators. 
Based on this observation they proposed  a new improved method of asymptotic dynamics which is  based on the asymptotic properties of matrix elements instead of operators and which also takes into account appropriate boundary conditions corresponding to the separation of particles at large distances.

The key observation is that the matrix element \cite{ANU05}
\begin{equation}
\langle \psi_{out}\vert H_{int}\vert  \psi_{in} \rangle 
\end{equation}
 is a time dependent complex number and, therefore, to investigate its asymptotic limit, one can use the method of stationary phase. Thus, if the above matrix element is given by\\
\begin{equation}
\langle \psi_{out}\vert H_{int}\vert  \psi_{in} \rangle=\int d\nu_i f(p_1)g(p_2)\cdots exp[-i\nu_i x^+] 
\end{equation}
then, according to the method of stationary phase, this integral approaches zero as $\vert x^+\vert\rightarrow \infty $ provided there is no point in the region of integration at which all first order partial derivatives of $\nu_i$  vanish.
The second criteria they suggest is to take into account the binding of particles at asymptotic limit. \\
In LFQED,
\begin{equation}
\nu_i=p^--k^--(p-k)^-
\end{equation}
The asymptotic region according to the KF method is given by $\nu_i=0$, whereas in the improved method, the asymptotic
region is defined by the following conditions,
\begin{equation}
\frac{\partial \nu_i}{\partial p_\perp}= \frac{\partial \nu_i}{\partial p^+}= \frac{\partial \nu_i}{\partial k_\perp}= \frac{\partial \nu_i}{\partial k^+}=0
\end{equation}

We have verified that for QED both the methods give the same asymptotic conditions but in case of theories with 4$-$point 
coupling the asymptotic regions obtained by KF method and the first criteria of improved method do not match. 
\section{Conclusion}

We have shown that the true IR divergences get cancelled when coherent state basis is used to calculate the matrix elements in lepton self-energy correction in light-front QED up to $O(e^4)$ in LF gauge as well as in Feynman gauge.
The cancellation of IR divergences between real and virtual processes is known to hold 
in equal time QED to all orders. It would be interesting to verify this all order cancellation in LFQED. 
The present work is an initial step in this direction. It is well known that IR divergences do not cancel in QCD in higher orders. This is related to the fact that the asymptotic states are bound states. Connection between asymptotic dynamics and IR divergences can possibly be exploited to construct an artificial potential that may be used in bound state calculations in LFQCD.
\begin{acknowledgements}
JM would like to thank Mumbai University and CICS for partial travel support for attending LC2013 and the organizer of LC2013 for their kind hospitality. AM would also like to thank DAE
BRNS, India for financial support during this project under the grant No. 2010/37P/47/BRNS.

\end{acknowledgements}

\begin{thebibliography}{3}
\bibitem{MUS91} D.\ Mustaki, S.\ Pinsky, J.\ Shigemitsu and K.G.\ Wilson, Phys.\ Rev.\ D {\bf 43}, 3411 (1991).
\bibitem{KUL70} P.P.\ Kulish and L.D.\ Faddeev, Theor. Math. Phys. {\bf 4}, 745
(1970).
\bibitem{ANU94} Anuradha\ Misra, Phys. Rev. D {\bf 50}, 4088 (1994).
\bibitem{ANU96} Anuradha\ Misra, Phys. Rev. D {\bf 53}, 5874 (1996).
\bibitem{ANU00} Anuradha\ Misra, Phys. Rev. D {\bf 62}, 125017 (2000).
\bibitem{JAI12} Jai\ D.\ More and Anuradha\ Misra, Phys.\ Rev.\ D {\bf 86}, 065037 (2012).
\bibitem{HILLER11} S.~S.~Chabysheva and J.~R.~Hiller, Phys.\ Rev.\ D {\bf 84}, 034001 (2011).
\bibitem{MUL98} R. Horan, M. Lavelle, and D. McMullan, Pramana 51, 317 (1998).
\bibitem{MUL99} R. Horan, M. Lavelle, and D. McMullan, Report No. PLY-MS-99-9, hep-th/9909044, (1999).
\bibitem{MUL00} R. Horan, M. Lavelle, and D. McMullan, hep-th /0002206 (2000).
\bibitem{ANU05} Anuradha\ Misra, Few-Body Systems 36, 201-204 (2005).
\end{thebibliography}


\end{document}